\documentclass[12pt]{iopart}
\usepackage{latexsym}

\def\Tr{{\rm Tr}}
\def\d={\buildrel \rm def \over =}
\def\ket#1{\mid~\!\!\!{#1}~\!\!\rangle}
\def\bra#1{\langle~\!\!{#1}~\!\!\!\mid}
\def\av#1{\langle~\!\!{#1}~\!\!
\rangle}

\begin{document}

\title[\bf Relative decoherence]{\bf Is
quantum decoherence reality or
appearance?}
\author{F Herbut\footnote[1]{E-mail:
fedorh@infosky.net}}
\address{Faculty of Physics, University of
Belgrade, POB 368, Belgrade 11001,
Yugoslavia and Serbian Academy of
Sciences and Arts, Knez Mihajlova 35,
11000 Belgrade}

\date{\today}

\begin{abstract}
\indent \small It has been
experimentally demonstrated that
quantum coherence can persist in
macroscopic phenomena [J.R. Friedman et
al.,Nature, 406 (2000) 43]. To face the
challenge of this new fact, in this
article QM in its standard form is
assumed to be extended by one beable
(hidden variable), i. e., a quantum
observable with always definite values
in nature (but usually only
statistically given in the quantum
state). Localization is taken as the
most plausible beable. The paradoxical
aspects of conventional QM take now a
different form. Suitably defining the
notion of "subject" fully within the QM
formalism, proving the quantum
conditional subsystem-state theorem,
and choosing the relative-decoherence
interpretation of QM, the paradoxes
formally disappear, leaving one with
decoherence relative to the definite
values of the beable; thus being only
appearance, not absolute reality in QM.
Relative to a different subject one has
perseverance of coherence. Hence, in
this approach it is claimed that {\it
decoherence and coherence , both exist
in reality}, but are not "seen" by the
same subject, and "subject" is, in this
interpretation, indispensable. The two
mentioned, apparently contradictory
phenomena, are in this way decoupled
from each other, contradiction is
avoided, and any one of the two can be
treated in the way that is usual in QM.
\end{abstract}

\maketitle

\normalsize \rm

\section{INTRODUCTION}

As it is well known, the linear
dynamical law of standard quantum
mechanics (QM) does not imply, in
general, definite measurement results
in disagreement with experience.
Namely, spontaneous evolution preserves
coherence, and definite measurement
results require decoherence. Therefore,
the basic question is the one in the
title of this article. It can be put
also as follows: Is decoherence  within
the quantum-mechanical formalism
absolute, i. e., a part of absolutely
objective reality, or, perhaps, it
appears only relative to a suitable
subject, i. e., isn't it actually only
appearance with restricted
quantum-mechanical reality?

Putting it in yet another way, the
question is whether coherence (or
superposition of states) perseveres in
macroscopic phenomena. Affirmative
experimental answers to this question
have been obtained recently \cite{Fri},
\cite{Orl99}.

Various interpretations of QM resolve
the mentioned fundamental difficulty by
going beyond QM in a number of
different manners \cite{PM}. The
Ghirardi-Rimini-Weber (GRW) theory
\cite{GRW} is, perhaps, the most
popular present-day approach to QM.

Hidden variables (HVs) \cite{B} attempt
to do this in another way with the
advantage that standard QM is left
unchanged (in contrast to what happens
in the GRW theory). Bell's theorem
\cite{Bell} claims that QM precludes
local hidden variables.

The well known fact that Bell's theorem
and its numerous proofs arose an
enormous interest in the literature
shows that local HVs were a great hope.
Nobody seems to like what appears, at
first sight, to be the only idea that
remains: nonlocal HVs.

Standard HV theories assume definite
values (in nature) for {\it all quantum
observables}. Within what remains
possible there is an alternative to
nonlocality: Assuming that only one
quantum observable has always definite
values in nature irrespective of the
quantum state. This observable is,
naturally, the position (or
localization) observable. We have the
case of definite-position extended QM
in the renowned approach of Bohm
\cite{dBB}, \cite{B}, and in some
speculations of Bell \cite{Bell'},
following whom we call this privileged
observable a {\it beable}.

In a recent article \cite{G} Ghirardi
has suggested an {\it experimentum
crucis} to distinguish between the GRW
theory and any approach that preserves
the whole of standard QM (like the
mentioned beable extension). Since the
suggested experiment should soon be
feasible, and since it is conceivable
that the outcome will be unfavorable
for the GRW theory, it is important to
consider a serious alternative. To
present such an alternative is the
basic aim of this article.

The approach presented below defines
"subject" in a simple way within QM,
and then treats it as indispensable. To
some extent this might fulfill the
expectation of Wilczek \cite{Wil00}
though he envisages "subject" as a
formidably complex notion.

It is shown below that in the suggested
one-beable-extended QM the mentioned
fundamental difficulty disappears, or,
to be more precise, it takes a form
that may be acceptable within QM.\\

\section{THE COHERENCE
WHICH IS THE STUMBLING BLOCK}

Let subsystem $1$ be the object on
which some dichotomic observable is
measured, and let subsystem $2$ be the
measuring instrument. Let, further,
$Q_2'$ and $Q_2''$, be two
orthocomplementary projectors for the
measuring apparatus such that their
occurrence (as events) means that the
"pointer observable" takes up the
respective "pointer positions" (meant
symbolically). Let, finally, {\it the
state into which the system $1+2$
evolves due to the measurement
interaction} be in terms of state
vectors (written as Dirac kets): $$
\ket{\Psi}_{12}=(w')^{1/2}\ket{\Psi}_{12}'+
(w'')^{1/2}\ket{\Psi}_{12}'',
\eqno{(1)}$$ where
$$\ket{\Psi}_{12}'\equiv
(w')^{-1/2}Q'_2 \ket{\Psi}_{12},\qquad
\ket{\Psi}_{12}'' \equiv
(w'')^{-1/2}Q''_2\ket{\Psi}_{12}
\eqno{(2a,b)} $$ (whether one,
actually, has $(1\otimes Q'_2)$ instead
of just $Q'_2$, is, of course, seen
from the context). The expansion
coefficients, which are assumed to be
positive, are given by the usual
expressions: $$w'\equiv
\bra{\Psi}_{12}Q'_2\ket{\Psi}_{12},
\qquad w''\equiv \bra{\Psi}_{12}Q''_2
\ket{\Psi}_{12}. \eqno{(2c,d)}$$

The {\it measurement paradox} consists
in the lack of definite measurement
results in (1), which has "inherited"
the, in general existing, {\it
coherence} in the initial state of the
object subsystem. However, if we assume
that QM is extended by {\it the
localization beable}, and that the two
"pointer positions" have a literal
meaning (they do not overlap spatially,
and their sum gives the entire space),
then {\it ipso facto}, as an immediate
consequence of the beable postulate,
for each {\it individual} $1+2$ system
the "pointer" will take up {\it one
position}, i. e., either $Q'_2$ or
$Q''_2$ will {\it occur in nature},
but, in our theory, this is {\it on the
beable level}.

There are now definite measurement
results. But the coherence of the two
terms (two results) in (1) persists,
and, as it is well known in QM, one
must apply the entire superposition to
the individual system. Hence, at first
sight, we have no less a paradox than
without the localization beable.

 Besides, having both coherence and
 decoherence at the
same time makes things confusing
because they seem to contradict each
other.

It is shown in this article that QM can
be so understood that only one
(coherence or decoherence) at a time is
taken into account. Thus, {\it
decoupling} one from the other, one can
make use of the quantum-mechanical
formalism in the usual way.

Returning to the above paradox, there
is, in addition, no way to decompose
the state
$\ket{\Psi}_{12}\bra{\Psi}_{12}$, which
is a {\it pure state in QM}, into the
two substates (empirically
subensembles) corresponding to the two
different results.

One has a similarly paradoxical
situation in the rest of the known
variations on the measurement paradox:
in the case of {\it "Schr\"{o}dinger's
cat"} \cite{S}, where, e. g., subsystem
$2$ is the "cat" and $1$ consists of
the rest of the deadly contraption; or
in {\it the Zurek environment theory}
\cite{Zu}, where $1$ is a classical
system, and $2$ is the environment.

In a further variation on the
measurement paradox, in the {\it
paradox of "Wigner's friend"} \cite{W}
subsystem $2$ includes the
consciousness of "Wigner's friend", who
looks at the "pointer" and sees its
"position". The coherent state (1) is
what "Wigner", or rather his
consciousness, "sees".

"Wigner" has the same difficulty as we
had in the measurement paradox: the
"friend" sees a definite result, and
"Wigner" himself knows, reading (1),
that both results are coherently
present in each individual case.\\

\section{THE EXPERIMENTAL SITUATION}

I have tacitly applied, what is called,
{\it QM of extended validity} in the
preceding section, viz. I have applied
QM to {\it macroscopic phenomena},
which involve millions (up to
$10^{23}$) microscopic constituents.
Otherwise, there is no paradox. It is a
well established fact that QM is fully
applicable to purely microscopic
phenomena.

Macroscopic objects and phenomena are
characterized by so-called {\it
macrorealism}, which, in our case
described by (1), claims that
$\ket{\Psi}_{12}$ has to be replaced by
the mixed state $$ \rho_{12}\equiv
w'\ket{\Psi}'_{12}\bra{\Psi}'_{12}+
w''\ket{\Psi}''_{12}\bra{\Psi}''_{12},
\eqno{(3)}$$ if the occurrences  of the
events $Q'_2$ and $Q''_2$ are {\it
classically distinguishable} because it
is an empirical fact that in such a
case there is no coherence. But is this
really so?

Many experimental and theoretical
physicists put in much effort to find
an answer, especially Leggett (see e.
g. \cite{Leg86} and the references
therein). Finally, an answer has been
obtained that has confirmed coherence
(against macrorealism) by two research
teams \cite{Fri} and \cite{Orl99} (cf
also \cite{Cho00}). This came after a
number of efforts with much progress in
demonstrating the macroscopic quantum
behaviour of various systems such as
superconductors, nanoscale magnets,
lasercooled trapped ions, photons in a
microwave cavity and $C_{60}$ molecules
(see references in \cite{Fri}).

In the two mentioned conclusive
experiments \cite{Fri}, \cite{Orl99}
$Q'_2$ and $Q''_2$ in (1) are SQUID
currents in the two opposite
directions, and the existence of (1)
has been ingeniously demonstrated.
Thus, millions of electrons can flow in
both directions simultaneously in a
coherent way. They do not behave
macrorealistically.

Still, one must be aware that a SQUID
is a macroscopic device with a quantum
phenomenon: the quantized magnetic
flux. (Phenomena of this type occur in
superconductors, superfluids etc.) We
are far away from an experimental
demonstration of coherent mixing of
classically distinguishable properties
in a purely classical object.
"Measuring instrument",
"Schr\"{o}dinger's cat", "Wigner's
friend" etc. are still in the realm of
speculations. But experimental
techniques are improving with
miraculous speed. Hence, I believe that
the approach of this article is
presented none too soon.

\section{THE "MONSTROUS" SUBENSEMBLES AND
THE CORRESPONDING QUANTUM MECHANICAL
SUBSYSTEM STATES}

Let us return to the first paradox
(that of measurement). Let us take the
empirical view and imagine the
(laboratory) ensemble described by
$\ket{\Psi}_{12}$. We denote it by $E$.
On account of the definite-localization
postulate, by which QM is extended in
the advocated approach, we can imagine
$E$ as consisting of two subensembles:
the one made up of those individual
systems on which $Q'_2$ occurs on the
beable level, and the one containing
the rest (on which $Q''_2$ occurs on
the same level). There is no way to
describe these subensembles in QM, i.
e., one cannot associate
quantum-mechanical states with them. We
call them "monstrous" subensembles, and
we denote them by $E'$ and $E''$
respectively.

We prove now that, nevertheless,
subsystem $1$ is described in the
"monstrous" subensembles by the
respective quantum-mechanical
conditional states (statistical
operators), which have the form as if
$Q_2'$ or $Q''_2$ occurred in ideal
quantum measurement on subsystem $2$
(on the quantum-mechanical level): $$
\rho'_1\equiv
(w')^{-1}\Tr_2\Big(\ket{\Psi}_{12}
\bra{\Psi}_{12}Q'_2\Big), \qquad
\rho''_1\equiv
(w'')^{-1}\Tr_2\ket{\Psi}_{12}
\bra{\Psi}_{12}Q''_2, \eqno{(4a,b)}$$
where "$\Tr_2$" denotes the partial
trace over subsystem $2$. The
probabilities of the two results, and,
at the same time, the statistical
weights of the two "monstrous"
subensembles $E'$ and $E''$ in $E$ , i.
e., $w'$ and $w''$, are given by (2c,d)
above.

 To prove  the stated {\it quantum
 conditional  subsystem-state theorem},
which is the basis of the approach
advocated in this article, we must give
{\it a precise definition of a beable},
which we assume to have a purely
discrete spectrum. (Since we are
concerned with position, one should
think of intervals, cf subsection
12.C.) Let us take a general quantum
state described by a statistical
operator $\rho$.

(i) Each particle always has a definite
value from the spectrum of the beable
(quantum-mechanical observable) in
nature irrespective of the quantum
state $\rho$. (If the quantum state
does not predict the mentioned beable
value, then the latter is called
"hidden").

(ii) When the beable observable is
measured, and the quantum state does
not predict a definite value of the
beable observable, then it gives on
each individual system the "hidden"
value of the beable. If the quantum
state does predict a sharp value, it is
the same as the value on the beable
level.

(iii) If the laboratory ensemble that
gives empirical representation to
$\rho$ has $N$ systems, then $N_i$ of
these have a given beable value, where
$\sum_iN_i=N$ (the summation is over
all beable values), and the relative
frequencies $N_i/N$ reproduce the
 probabilities predicted by QM in the usual
way: $$N\rightarrow \infty ,\quad
\Rightarrow \quad N_i/N\Rightarrow w_i=
\Tr\rho Q_i, \eqno{(5)} $$ where $Q_i$
are the characteristic projectors of
the beable observable, and $w_i$ are
the statistical weights of the
"monstrous" subensembles in the entire
enseble that gives empirical
representation to $\rho$. (This is a
generalization of $w'$ and $w''$, cf
(2c,d).)

A proof of this theorem is given in the
Appendix. In previous work \cite{H} the
second requirement in the above
definition of a beable observable was
put more broadly, and the quantum
 conditional  subsystem-state theorem
was proved in a more complicated way.

Thus, as far as subsystem $1$ is
concerned, the "monstrous" subensembles
$E'$ and $E''$ are seen to be described
by statistical operators, and thus,
they are normal quantum-mechanical
states (with no monstrosity
whatsoever). The subensembles $E'$ and
$E''$ are still monstrous regarding
subsystem $2$ and with respect to the
whole (in general, entangled) $1+2$
system.

Now it is clear that, if we want to rid
ourselves of the monstrosities, we must
find a way to restrict consistent
quantum-mechanical description to
subsystem $1$, and, nevertheless, we
must be able to recover the description
of subsystem $2$ and of the entire
system.\\

\section{SHIFT OF THE SPLIT AND
CONVERSION OF THE ENVIRONMENT INTO
SUBJECT}

The {\it object} ($O$) of
quantum-mechanical description is, as a
rule, a restricted part of the
universe, and the rest, usually called
the environment, is left out. In the
decoupling of decoherence from
coherence, that I am striving to
expound, I want only the former to
belong to the object. But I do not want
coherence to be quite left out. It
should be present in the omitted
environment, converting it into, what I
call, the {\it subject} ($S$) (playing
the role of a kind of "observer" that
is defined strictly quantum
mechanically). Between the object and
the subject there is an imaginary {\it
cut} ($/$); altogether we have a {\it
split}: $O/S$. If the environment is
not yet converted into subject, we
write $O/\dots$.

The cut is due to a subjective choice
on our part, and it can be {\it
shifted} to broaden the object. This is
important for the introduction of the
beable observable in order to convert
the environment into subject.
Afterwards, the cut is shifted back.

Let us elaborate this on the mentioned
paradox of measurement theory.\\

\section{DEFINITE BUT RELATIVE
MEASUREMENT RESULTS}

Relation (1) applies to the composite
system object plus measuring apparatus
at the end of the measurement
interaction. Calling this system $1+2$,
we can start with the split $O/\dots
\equiv 1/(2+$the rest of the universe).
To convert the environment, which is
now ($2+$the rest of the universe),
into a subject, we shift the cut
broadening the object, i. e., we go
over to $O/\dots \equiv (1+2)/$(the
rest). Then the {\it localization}
beable observable can be normally
introduced as a suitable
second-subsystem observable (with only
two characteristic projectors $Q'_2$
and $Q''_2$ - corresponding to two
intervals on the pointer scale - in our
simplified discussion). We can then
shift the cut back to obtain a
description of subsystem $1$ with {\it
the environment converted into a
subject}: $O/S\equiv 1/(2+$the rest).

Now the desired decoupling of
decoherence from coherence is achieved.
At the end of measurement we have
subsystem $1$ as our object of
quantum-mechanical description. We have
the decoherence: $$
\rho_1=w'\rho'_1+w''\rho''_1,
\eqno{(6)}$$ where $\rho_1\equiv
\Tr_2\ket{\Psi}_{12} \bra{\Psi}_{12}$
is the subsystem state decomposed into
the two quantum conditional
subsystem-states  defined in (4a,b)
(the respective first-subsystem states
of the composite "monstrous"
subensembles $E'$ and $E''$).

We have now {\it relative decoherence},
namely, for each individual-system case
the first-subsystem state is either
$\rho'_1$ or $\rho''_1$ (but not both,
i. e., not $\rho_1$), and it is a state
{\it relative to the "pointer
position"}, i. e., relative to the
occurrence of $Q'_2$ or $Q''_2$ on the
beable level. In this manner one has
definite results in a
quantum-mechanical description.

There is nothing wrong with the {\it
objectivity} or {\it realness of the
decoherence} on the quantum-mechanical
level. But this objectivity is not in
the sense of absolute observer
independence, as one expects in the
case of {\it absolute decoherence}
(which is much favored in the
literature \cite{PM}, \cite{GRW}).

 The decoherence is "real"
for anyone who "looks" at the
"pointer", or to whom this observer
might communicate the result of
measurement. The mentioned "looking" or
the mentioned communication amounts to
an interaction that results in a new
composite pure state of the type (1),
in which subsystem $2$ now contains,
besides the "pointer positions", also
the consciousness of the first and
possibly that of other observers. We
have now a more intricate composite
subject. In case there are two or more
human observers, they all see the same
measurement result.

In a theory with subjects, "reality" or
"objectiveness" is recognized by
intersubjective agreement. Naturally,
the expounded relative decoherence does
have the attributes of reality because
there is sufficient intersubjective
agreement. Though, there is no
universal intersubjective agreement as
one would have in the case of absolute
decoherence. Thus, one might say that
decoherence is only appearance within
the quantum-mechanical formalism.

Returning to the case prior to
communication, when subsystem $2$
contained only the "pointer positions",
we can say that the {\it coherence} has
not disappeared. We can take another
measurement apparatus as subsystem $3$,
and define its "pointer observable" as
a beable observable. For simplicity, we
assume that we have again localization
intervals as the "pointer positions".
(For a more general case, cf subsection
12.G.)

We make the shift to the split $O/\dots
\equiv (1+2+3)$/(the rest), introduce
the new beable observable, and shift
back to $O/S\equiv (1+2)/(3+$the rest).
The environment is again converted into
a subject, and the second apparatus
(possibly handled by a second observer,
all included in subsystem $3$) begins
the second measurement on the coherent
state given by (1).

The {\it coherence} in (1) is {\it
real}, as the second observer can
easily ascertain by performing a
measurement of the coincidence
occurrence of two suitably chosen
subsystem events (projectors):
$P_1P_2$. The suitability of the
subsystem projectors consists in their
ability to give a different probability
in the pure state $\ket{\Psi}_{12}$
given by (1) than in the corresponding
decoherent quantum state $\rho_{12}$
(cf (3)). Since the former probability
is $$
\bra{\Psi}_{12}P_1P_2\ket{\Psi}_{12}=
w'\bra{\Psi}_{12}'P_1P_2\ket{\Psi}_{12}'
+w''\bra{\Psi}_{12}''P_1P_2
\ket{\Psi}_{12}''+$$ $$
(w')^{1/2}(w'')^{1/2}\bra{\Psi}_{12}'
P_1P_2\ket{\Psi}_{12}''+(w'')^{1/2}
(w')^{1/2}\bra{\Psi}_{12}''P_1P_2
\ket{\Psi}_{12}', $$ and the latter, i.
e., $\Tr\rho_{12} P_1P_2$, equals the
sum of the first two terms above,
"suitability" amounts to the last two
terms giving a nonzero number. Such a
choice of the projectors is, in
principle, possible, and therefore {\it
the reality of the coherence} in (1) is
an experimentally establishable fact.

By decoupling we have achieved that
decoherence and coherence now {\it do
not contradict each other}. Namely,
they are relative to different
observers: the decoherence is relative
to the first measuring apparatus (and,
possibly, the first observer if it is
included in subsystem 2) at the end of
the first measurement, and the
coherence is relative to the second
measuring apparatus (and, possibly, the
second observer) at the beginning of
the second measurement.

The important point to notice is that,
like in my previous article on the
bubble chamber \cite{FHM}, I propose an
interpretation of QM, which may be
called {\it the relative-decoherence
interpretation}, in which
quantum-mechanical prediction {\it
always presupposes} prior convertion of
the environment into a subject, i. e.,
introduction of a suitable beable
observable defined in a subsystem of
the environment.

This is hardly a new interpretation
having in mind that Niels Bohr insisted
on the importance of observer and on
its inseparability from object in QM,
as it is well known. I only propose a
very concrete way of doing this by
shrinking Bohr's well known idea of the
role of a classical measuring apparatus
to a simple beable observable (with a
discrete spectrum), which is mainly
localization in a perfectly classical
sense. This approach stands opposed to
the approaches without subject and with
absolute decoherence, especially to the
GRW theory \cite{GRW}.\\

\section{THE ROLE OF ENVIRONMENT
ACCORDING TO ZUREK}

According to the Zurek theory
\cite{Zu}, interaction with the
environment can lead to absolute
decoherence on account of his claim
that it is position-measurement-like,
and thus explanation is obtained why
classical objects do have definite
positions (as opposed to superpositions
of them).

Nevertheless, it is hard not to agree
with the obvious criticism that goes as
follows: If one takes a sufficiently
large system consisting of the
classical objects plus environment
(possibly the whole universe), then the
interaction will, if it is precisely
position-measurement-like, give a
composite state vector as in (1), where
$1$ are the classical objects and $2$
is the environment; where, further, not
only $Q'_2$ and $Q''_2$ represent
different distributions of matter, but
also the corresponding first-subsystem
states $\rho'_1$ and $\rho_1''$ imply
nonoverlapping different definite
positions with certainty. But there is
no way to obtain absolute decoherence,
because the rules of QM require that
each individual $1+2$ system be
described by $\ket{\Psi}_{12}$ (cf (1))
(cf also D'Espagnat's discussion
\cite{d'E} of improper mixtures as
$\rho_1$ in (6) is).

On the other hand, if one assumes that
position of every particle is a beable,
then so is the matter distribution in
the environment (subsystem $2$), and,
utilizing it, we can convert the
environment into, what I have suggested
to call a "subject".

Then we do have absolute decoherence of
the distinct positions of subsystem
$1$, but only outside QM, via the
"monstrous" subensembles $E'$ and $E''$
respectively (cf section 3). To bring
this into the formalism of QM, one may
apply a procedure as in the preceding
section and obtain {\it relative
decoherence} in $\rho'_1$ and
$\rho''_1$, but this time relative to
the different matter distributions in
the environment.

One still has coherence in the larger
system $1+2$, but it is irrelevant for
the classical objects (subsystem
$1$).\\

\section{"SCHR\"{O}DINGER'S CAT"}

Let now $\ket{\Psi}_{12}$ given by (1)
describe the end of the thought
experiment with Schr\"{o}dinger's cat
\cite{S}, where $Q'_2$ and $Q''_2$ are
associated with the dead and the live
cat respectively. In full analogy with
the case of end of measurement
interaction (section 6),
"Schr\"{o}dinger" himself may be
subsystem $3$, who in the
individual-system case "sees" a
superposition of the dead and live cat
described by (1). In spite of this, it
can be seen in two ways that the
individual cat is either dead or alive:

    (i) Let subsystem $2$ be the cat,
and let one assume, further,  that the
consciousness of the cat is a beable,
and $Q'_2$ and $Q''_2$ are projectors
meaning "dead" and "live" respectively.
Then the cat can be the subject, and
all inanimate parts of the deadly
contraption, i. e., subsystem $1$, be
the object with state decomposition (6)
expressing {\it relative decoherence}
of the decayed atom (and broken ampula)
and the undecayed atom (and whole
ampula). Again, of course, this
quantum-mechanical decoherence takes
place relative to the contents of the
consciousness of the cat, or  rather
relative to "extinguished
consciousness" and "active
consciousness" respectively.

(ii) One can introduce the position
beable observable for the material
particles, and take for subject the
second subsystem with the distribution
of matter in the ampula. Then the dead
and live states of the first subsystem,
the object, which is now the cat, are
decoherent relative to the distinct
distribution of matter in the ampula in
case of decay and nondecay of the atom
respectively. This case is symmetric to
the previous one: $Q'_2$ and $Q''_2$
express the mentioned distinct
distributions of matter, and (6)
applies to the ensemble that is a
mixture of dead and live individual
cats.

In the advocated approach, one must
decide on one of the two preceding
alternatives for choice of object and
subject. One may change over from one
to the other, but one cannot have both
simultaneously.

    Though the choice of subject and
object is arbitrary, once made, one
must stick to it in one discussion.
This is similar to the case of tensor
product in treating composition of
subsystems in the formalism of QM: one
can change over to a different order of
the tensor factor state spaces (the
change is an isomorphism), but one must
stick to one order for the whole of an
argument.

Returning to any of the two
possibilities of discussing
Schr\"{o}dinger's cat, one must
emphasize that "Schr\"{o}dinger"
himself "sees" coherence, because his
object is $(1+2)$.

We see again that introduction of a
beable brings in coexistence of
decoherence (in the "monstrous"
subensembles on the beable level) and
coherence (on the QM level). To treat
both within QM, one can decouple them
from one another by a suitable choice
of subject and object (as
illustrated).\\

\section{"WIGNER'S FRIEND"}

Let us take as a final illustration the
mentioned case of {\it "Wigner's
friend"}. Having system $1+2$ in mind,
we can start with the split $O/\dots
\equiv (1+2)/$(the rest). To introduce
a suitable subject, we can assume, with
von Neumann, that human consciousness
is always definite in the case when an
observer is looking, i. e., that it is
a beable (though von Neumann might not
have agreed with treating his idea in
this way).

This beable is, actually, an empirical
fact (we know that it is so), put as a
postulate into extended QM.

Thus, let us take the beable observable
which is the consciousness of "Wigner's
friend". It is a second-subsystem
observable with only two characteristic
projectors $Q'_2$ and $Q''_2$ in our
simplified discussion. We can then
shift the cut to narrow down the object
(and to get rid of the coherence in
it), i. e., we may go over to the split
$O/S\equiv 1/(2+$the rest) (with a
well-defined subject).

Now the desired decoupling of
decoherence from coherence is
completed. The object of
quantum-mechanical description of
"Wigner's friend" is subsystem $1$. The
"friend" "sees" the decoherence
expressed in relation (6).

We have now {\it relative decoherence},
viz., for each individual-system case
the first-subsystem state is either
$\rho'_1$ or $\rho''_1$, and it is a
state {\it relative to the contents of
the consciousness} of "Wigner's
friend", who sees a definite result.

We can say that the {\it coherence} has
not disappeared. We can take "Wigner"'s
mind as subsystem $3$, and define
"Wigner"'s consciousness as a beable
observable. We make the shift to
$O/\dots \equiv (1+2+3)$/(the rest),
introduce the new beable observable,
and shift back to $O/S\equiv
(1+2)/(3+$the rest). The environment is
again converted into a subject, and
"Wigner" "sees" the coherent state
given by (1).

By decoupling decoherence from
coherence, we have achieved that they
do not contradict each other. This is
so because they are relative to
different observers: the first is
relative to "Wigner's friend", and the
second is relative to "Wigner".\\

\section{RELATION TO EVERETT}

The definition of the conditional
first-subsystem state (4a) even with a
general composite-system state
$\rho_{12}$ replacing
$\ket{\Psi}_{12}\bra{\Psi}_{12}$
 and with $Q'_2$ as an arbitrary
second-subsystem event (projector), is
a part of standard QM formalism. Its
physical meaning is the conditional
state in the case of ideal measurement
of $Q'_2$ in $\rho_{12}$. This is well
known, because the standard L\"{u}ders
formula, which describes change of
state in ideal measurement, converts
$\rho_{12}$ in the ideal occurrence of
$Q'_2$ into
$(w')^{-1}Q'_2\rho_{12}Q'_2$, and
partial trace of this is the RHS of
(4a).

That $\rho_1'$ has the physical meaning
of the conditional state in {\it any}
occurrence of $Q'_2$ is not well known,
but true \cite{HV}. As it has been
demonstrated in this article, this
physical meaning extends even to the
case when $Q'_2$ occurs on the beable
level. I have called $\rho'_1$ {\it the
relative state}, meaning "the state
with relation to" the beable occurrence
of $Q'_2$.

My "relative state" formally
generalizes that of Everett \cite{Ev}.
Namely, when $\rho_{12}$ is a pure
state $\rho_{12}\equiv
\ket{\Psi}_{12}\bra{\Psi}_{12}$, and
$Q'_2$ is an elementary projector (a
quantum logical atom) in the state
space of the second subsystem, i. e.,
$Q'_2\equiv \ket{\phi}_2\bra{\phi}_2$,
my relative state reduces to Everett's.
One can choose a second-subsystem
complete orthonormal basis with
$\ket{\phi}_2$ as one of the basis
vectors, expand $\ket{\Psi}_{12}$ in
this basis, and then one has Everett's
formal context in which the conditional
state $\rho_1'$, which is now also a
pure state, appears. This is what he
called "relative state".

Since Everett did not introduce
beables, and since the choice of the
mentioned basis was quite arbitrary,
his relative state was physically ill
founded. I believe that he himself
realized this and that his construction
of the "branching of the universe" was
seeking a way out of this. But, I agree
with Shimony \cite{Shi} that Everett's
"branching" is only a semantic
innovation, not a new idea, and
therefore also Everett's "branching of
the universe" is not less ill founded
conceptually. The universe is hard to
be thought of as "branching" in a
noncountable infinity of ways at the
same time (because this is the amount
of nonuniqueness in the choice of
Everett's basis).

Anyway, I consider Everett's article as
an important step towards the {\it
relative decoherence interpretation} of
QM advocated in this article. (I have
learnt a good deal from Everett.)\\

\section{RELATION TO THE THEORY OF
GHIRARDI, RIMINI AND WEBER}

The mainstream of interpretative
thought on the foundations of QM goes,
as well known, towards {\it absolute
decoherence} within QM. In my
simplified presentation in this article
this would amount to transition from
the pure composite system state
$\ket{\Psi}_{12}$ given by (1) to the
mixed composite state given by $$
\rho_{12}\equiv Q'_2
\ket{\Psi}_{12}\bra{\Psi}_{12}Q'_2+
Q''_2\ket{\Psi}_{12}\bra{\Psi}_{12}Q''_2.$$

Since this cannot be achieved by the
use of purely QM ideas, one always adds
an extra-quantum-mechanical
stipulation.

The Ghirardi-Rimini-Weber (GRW) theory
\cite{GRW} seems to me to be the climax
of absolute-decoherence endeavors. As
such it has a strong appeal to many
leading foundationally and
realistically minded physicists, who
try to comprehend what QM is really
about.

Basicly, I have no objection to the GRW
theory. If one can interpret QM in an
objective way without "observer" (cf
Bell's criticism in \cite{Bell''}), one
should do so. If it were a question of
taste, perhaps I myself would be in
favor of the GRW alternative, and
against my advocated
relative-decoherence interpretation.
But this is not a matter of taste. It
is a question of fact if the coherence
persists in the macroscopic phenomena
or not.

There is much theoretical and
experimental research done these days
determined to discover the answer to
this question  (see section 3). I am
sure that in a few years we will know
with certainty if the coherence in
measurement (or measurementlike
processes like the interaction with the
environment in the Zurek theory)
persists or not in macroscopic
phenomena. If it does not, i. e., if
{\it decoherence is fact} (in an
absolute sense), one can forget about
the  relative-decoherence
interpretation advocated in this
article. But if coherence does persist,
if it is {\it reality}, then {\it
decoherence is only appearance} (in
QM). In this case, one will hardly find
a simpler way than the expounded
proposal of relative decoherence to
express the mentioned persistence of
coherence not going outside the
formalism of standard QM .\\

\section{CONCLUDING REMARKS}

 \bf A) Why one hidden variable?
\rm The definite-position extension of
QM advocated in this article is a {\it
minimal extension}. Quantum mechanics
allows one to do such an extension with
any one observable (or any set of
compatible observables) in a natural
way and without no-go troubles. The
advocated theory is {\it sufficient}
for our purposes. It does not preclude
a wider consistent extension. After
all, I have myself added a second
beable in the form of the consciousness
of "Schr\"{o}dinger's cat" or of
"Wigner's friend", cf also subsection
12.G. (It is likely that consciousness
as an observable is compatible with the
localization observable, so that,
formally, we have extension with one
observable, which is a set of two.)
\\

 \bf B) Why the position observable?
\rm Space (the spectrum of the position
observable) does, no doubt, occupy a
{\it unique place in physics}. We have
relativity and kinematics (both
nonrelativistic and relativistic) in
space. The spectra of other observables
do not have so much physical
significance.\\

\bf C) The position observable has a
purely continuous spectrum. \rm The
definite position stipulation, of
course, assumes that each particle (or
pointlike part of a particle when the
latter has a finite volume) has a
definite position, i. e., a point from
the continuous spectrum of the position
observable. But, as well known, no
measurement gives as a result the point
of the continuous spectrum. Hence, the
subject observable is naturally defined
as one with a purely discrete spectrum.
Any breaking up the real axis into
nonoverlapping intervals gives,
essentially, such an observable, and
the particle (or part of it) belongs to
precisely one of the intervals. (We had
only two intervals in our simplified
approach.)\\

\bf D) Spatial nonlocality. \rm One
wonders if definite-position QM is
compatible with the known fact of the
existence of spatial nonlocality like
the one in Young's two-slit
interference \cite{Hint} or in the
Mach-Zehnder interferometer. (Both have
been performed with massive particles
\cite{Su82}, \cite{Ze88}.)

    If we want to imagine a "physical
mechanism" in these experiments, then
standard (unextended) QM offers that of
{\it a delocalized particle} that
"passes both slits" or "goes through
both branches" of the interferometer,
and then, when the two latent
 or potential "paths"
meet, they interfere.

Adherents to the definite-position
extension of QM reject such "potential
paths" as physical nonsense. To their
minds the "physical mechanism" is
necessarily based on real, though
unknown, paths: the individual particle
goes through one slit or through one
branch. But then a nontrivial price has
to be paid for an explanation of the
interference. There seems to be no
other way than to reinstall some kind
of \it action at a distance \rm in
physics: the particle must "feel" if
the other slit is open or if the other
branch can be traversed or not. (See,
e. g., the fascinating experiments on
interactionfree measurement
\cite{Kw95}, \cite{Kr96}, and
\cite{Kw96}.)

One way to do this is in the manner of
the de Broglie-Bohm school of thought:
through nonlocal potentials depending
on the wave function of the particle
\cite{B}. This may seem not more
"palatable" than the above mentioned
delocalization in the way of potential
paths. But spatial nonlocality (on the
one-particle level!) is a fact that
must be faced either in a kinematical
way as in unextended QM through
delocalization, i. e., by being
"smeared out in space", or in a
dynamical way as, e. g., in the
mentioned de Broglie-Bohm theory. (Or
in a third way if anybody can think of
it.)\\

 \bf E) The Role of Classical Physics.
 \rm Niels Bohr, repeatedly
emphasized that {\it the validity of
classical physics must be assumed on
the same level as that of QM} in order
to be able to interpret measurement
results and hence QM itself.

{\it The definite-position stipulation
can be viewed as a realization of this
idea of Bohr}: It introduces into the
extended theory {\it a sufficient
amount of classical ideas}, and, on the
other hand, in contrast to the
Copenhagen approach \cite{St}, {\it
these ideas are introduced in a
perfectly natural way as far as QM is
concerned}. Namely, any density matrix
gives a probability measure on the
spectrum of any observable. In view of
the fact that measurements are made on
ensembles of individual systems, the
natural way to interpret such a
probability measure is the
definite-value stipulation of the
observable at issue on the
individual-system level. This is called
"ignorance interpretation" of the
probability in QM. If all observables
are encompassed, one has a standard
hidden-variables theory (cf also
subsection G) below).\\

 \bf F) Objectivity of QM. \rm
The advocated approach assumes extended
QM (encompassing measuring instruments,
"cats" and "friends"). Then in nature
both coherence and decoherence are
fully objective. In QM decoherence (in
an absolute and universal or fully
subject-independent sense) is only {\it
appearance}; its reality is restricted
to suitably chosen subjects, i. e., it
is relative. Also coherence in QM is
observer dependent in the advocated
relative-decoherence interpretation.\\

{\bf G) What about nonspatial "pointer
observables" in measurement?} \rm It is
known that the different "pointer
positions" can be distinct values of
some observable that is not even
compatible with position and that these
"positions" of the "pointer observable"
can take place at one and the same
location. To cover this case, the
theory should be further extended by
the "pointer observable" in question,
so that, at the end of the measurement
interaction, we have a two-
hidden-variables extension. This is
possible, as one easily concludes from
the fact that QM allows extension by
{\it all} hidden variables, i. e., all
observables can be considered to have
definite values from their spectra at
the price of introducing contextuality
(\cite{B}, \cite{Dj91}, \cite{Vuj94}).
As a consequence of contextuality, the
definite value is attributed to
observable plus context. In our  case,
we have only two contexts: the one
defined by the position, and the one
defined by the "pointer observable",
both in the sense of minimal
measurement \cite{He}. ( We forget
about all the rest of hidden
variables.) These two will raise no
intuitive paradox of nonlocality.

    One wonders how the definite values of
the "pointer observable" may come
about, since, in our
one-hidden-variable theory, these
values are not always definite. There
seem to be two possibilities:

    (a) The very dynamics on the
subquantum level that brings about some
classical systems, like the measuring
instrument, may be responsible for the
definite values of the "pointer
observable".

    (b) It may be that some measurement
interactions give rise to them on the
beable level.\\

\bf H) Subquantum dynamical law. \rm If
a particle has a definite position, as
assumed in this approach, one must find
a law how it changes spontaneously in
time (a subquantum dynamical law). The
theory of Bohm et al \cite{dBB}
\cite{B} does give an answer to this
question. But there are also other
ideas \cite{Bell'}.\\

\bf I) Relation to the universe. \rm In
conventional QM \cite{St} it is hard to
make sense of the idea of a wave
function of the universe. In the
advocated relative-decoherence
interpretation of QM all reality lies
in the mutual statistical correlations
between the different parts of the
universe, and the mentioned wave
function need not have more meaning
than that.

In conventional QM one may think of
obtaining the quantum state
(statistical operator) of a part of the
universe by tracing out all other
subsustems in the mentioned wave
function (written as a statistical
operator). In the relative-decoherence
interpretation, one must allow for the
definition of the subject observable in
the environment (as explained), and
hence it cannot be irrevocably traced
out.\\

 \bf J) From conditional
probabilities to first-person
description. \rm Careful analysis
reveals that one requires collapse only
when a second measurement is to be
performed on the basis of a fixed
result in the first measurement.
Adherents to the no-collapse approach
repeatedly point out that one can do
without the idea of collapse by
utilizing conditional probabilities.

The weak point of this view is that
this approach lacks the very idea of
occurrence of an event, and hence there
is no way to understand what the
"condition" of the conditional
probability is. Besides this conceptual
incompleteness, there is also the fact
that this approach is a purely
third-person description, as if God
were viewing what is going on in this
world. The approach does not allow one
to introduce a first-person
description, and it is a
phenomenological fact that, in
principle, I can do the experiment and
it is natural that I want a
first-person description. The
relative-decoherence approach advocated
in this article provides one both with
the idea of occurrence (possession of a
certain position on the beable level),
and with a first-person description.\\

\bf K) Relation to my previous
"relative-collapse interpretation" \rm
\cite{FHM}. "Relative collapse" and
"relative decoherence" are, of course,
one and the same thing. They differ
only in emphasis whether it is on the
definite-result subensemble (so-called
selective measurement) or on the entire
ensemble (so-called nonselective
measurement). So I am advocating one
and the same interpretation. But there
is a difference between the position I
took in my preceding article \cite{FHM}
and that in the present one. Namely, in
the former, the relative-collapse
postulate, as I called it, assumed that
each individual quantum system
possesses a definite value of the
subject observable, which was any
chosen fixed observable with a purely
discrete spectrum. The assumed definite
values were a kind of self-collapse.
This was a part of the definition of
the subject. It was difficult to accept
all this. In the present paper there is
mainly one subject observable: the
position observable. The mysterious
self-collapse is replaced by the
stipulation that each particle always
has a definite position in nature. This
is easier to accept.\\

\noindent {\bf ACKNOWLEDGEMENT}\\

\noindent The author is indebted to
Prof. Anton Zeilinger for his
invitation. Thanks are due also for the
financial support and hospitality of
the "Erwin Schr\"{o}dinger" Institute
in Vienna, where part of this work was
done.

\section{APPENDIX}

{\it Proof of the  the quantum
 conditional  subsystem-state
 theorem}.

We return to (1), and assume that
$Q'_2$ and $Q''_2$ are the two
characteristic projectors of our
dichotomic beable observable for
subsystem $2$. Let, further, $C_1$ be
an arbitrary first-subsystem observable
with a purely discrete spectrum , and
let $\av{C_1\otimes 1}_{E'}$ be the
average of $C_1\otimes 1$ in {\it the
"monstrous" subensemble} $E'$,
determined, in principle, by the
occurrence of $Q'_2$ on the beable
level. (I do not assume that the
average is a finite number.) Finally,
let the $N'$ individual systems in the
subensemble $E'$ be ordered in some
arbitrary but fixed way and enumerated
by $j=1,2,\dots ,N'$. Let then
measurement of $C_1$ produce the
results $\{c_j:j=1,2,\dots ,N'\}$. One
can write: $$ \av{C_1\otimes 1}_{E'}=
\sum_{j=1}^{N'}c_j/N'=(N/N')\sum_jc_j/N
\rightarrow (w')^{-1}
\Tr_{12}\ket{\Psi}_{12}\bra{\Psi}_{12}
(C_1\otimes Q'_2)=\Tr_1C_1\rho'_1,$$

$$ \mbox{\rm when} \quad N\rightarrow
\infty $$ (cf (4a)). (I have here
tacitly assumed that the ideal
quantum-mechanical occurrence of $Q'_2$
does not change any of the values $c_j$
on subsystem $1$. It only reveals the
hidden beable value of $Q'_2$. This is,
actually, a tacit locality assumption
on the beable observable.)

 The proof for the other
value of the beable (where $N'',\enskip
E''\enskip Q''_2$ appear) runs, of
course, symmetrically. This ends the
proof.\hfill $\Box$\\

\end{document}